\documentclass[sigconf]{acmart}

\usepackage{algorithm}
\usepackage{algorithmicx}
\usepackage{algpseudocode}
\usepackage{caption}
\usepackage{subcaption}
\usepackage{soul}

\newcommand*\rfrac[2]{{}^{#1}\!/_{#2}}

\AtBeginDocument{%
  \providecommand\BibTeX{{%
    \normalfont B\kern-0.5em{\scshape i\kern-0.25em b}\kern-0.8em\TeX}}}

\setcopyright{acmcopyright}
\copyrightyear{2018}
\acmYear{2018}
\acmDOI{XXXXXXX.XXXXXXX}

\acmConference[Conference acronym 'XX]{conference title}{June 03--05,
  2018}{Woodstock, NY}
\acmPrice{15.00}
\acmISBN{978-1-4503-XXXX-X/18/06}

\begin{document}

\title{CrowDC: A Divide-and-Conquer Approach for Paired Comparisons in Crowdsourcing}


\author{Ming-Hung Wang, Chia-Yuan Zhang, Jia-Ru Song}
\email{tonymhwang@ccu.edu.tw, {kech880604,sjr}@csie.io}
\affiliation{%
  \institution{Department of Computer Science and Information Engineering}
  \streetaddress{}
  \city{}
  \country{National Chung Cheng University, Taiwan}
}

\renewcommand{\shortauthors}{Wang, et al.}

\begin{abstract}
Ranking a set of samples based on subjectivity, such as the experience quality of streaming video or the happiness of images, has been a typical crowdsourcing task. Numerous studies have employed paired comparison analysis to solve challenges since it reduces the workload for participants by allowing them to select a single solution. Nonetheless, to thoroughly compare all target combinations, the number of tasks increases quadratically. This paper presents ``CrowDC'', a divide-and-conquer algorithm for paired comparisons. Simulation results show that when ranking more than 100 items, CrowDC can reduce 40-50\% in the number of tasks while maintaining 90-95\% accuracy compared to the baseline approach.

\end{abstract}




\maketitle

\section{Introduction}


Paired comparison analysis is a common method used in crowdsourcing to rank samples based on subjective criteria. It involves presenting workers with pairs of samples and asking them to choose the better option. This method has been used in psychology, economics, and engineering, to assess subjective qualities such as the quality of streaming video and aesthetics of graphic design \cite{wu2013crowdsourcing,wu2021learning}.

However, if the number of items is high, the number of tasks required to compare all target combinations will grow quadratically, which might burden workers and require more time and costs to complete all tasks. To overcome this challenge, some researchers have proposed alternate approaches for ranking subjective samples, such as using aggregation or learning from a few human labeling to improve the labeling quality or reduce the comparisons required \cite{gleich2011rank,shah2016stochastically}. However, if partial ranks from crowds are biased and unfair, they would decrease the reliability of the results.

Thus, this work aims to reduce the number of tasks required by refining the workflow. We develop a divide-and-conquer-based algorithm, ``CrowDC,'' reducing the number of tasks required for paired comparison analysis while maintaining high accuracy. First, we ``Divide'' the to-be-evaluated sample combinations into multiple groups. The proposed algorithm utilizes the Bradley-Terry-Luce (BTL) approach to compare all sample combinations for each group and calculate their estimation scores. To find inter-group relationships, the proposed algorithm then compares several pivots from each group. Finally, the proposed algorithm aligns and merges all groupings of samples, a process referred to as ``Conquer.''

To validate our design, we conducted simulations with various parameters for group size, number of pivots, the accuracy of workers, etc. In some scenarios, our method can reduce up to 50\% in the number of tasks while maintaining 90-95\% accuracy compared to the BTL approach when ranking 100 items or more. Hence, our proposed algorithm is particularly beneficial when human labeling costs are high, enabling a more efficient and cost-effective method for ranking subjective paired comparison samples.

\section{Related Works}

One of the advantages of paired comparison analysis is that it is straightforward for workers to understand and complete, reducing workload and improving process efficiency~\cite{bradley1952rank}. Hence, this method has been implemented in numerous crowdsourcing applications, such as measuring the quality of experience~\cite{wu2013crowdsourcing}, image quality~\cite{xu2012online}, and the aesthetic quality of chart layout~\cite{wu2021learning}.

Nevertheless, paired comparison analysis has its limits. Specifically, ranking numerous samples can be time-consuming and costly. For example, if there are $s$ items to be evaluated, and workers are asked to compare two items at a time, the required tasks to compare all combinations would be $\binom{s}{2}$. Because it relies on subjective judgments, it may be affected by individual biases and perceptions, impacting reliability and validity~\cite {draws2021checklist,duan2022influences}.

The Bradley-Terry-Luce (BTL) approach is a common statistical model used to predict the likelihood that one item is preferred over another when using paired comparisons to rank samples~\cite{bradley1952rank,luce2012individual}. The main benefit of the BTL approach is its capability to be fitted to data and reveal differences in item choices. However, the BTL approach has a major drawback: it may not perform effectively when comparing many items with a small number of comparison outcomes. Also, the BTL approach requires that all obtained items be compared, which is time-consuming and lowers experimental efficacy due to its computing complexity. 


To overcome these limitations, we propose a divide-and-conquer-based algorithm, ``CrowDC,'' aiming to significantly reduce the cost of tasks while maintaining a high accuracy comparable to that achieved through paired comparison analysis.

\section{Methodology}

In this section, we introduce two methods for paired comparison: the Bradley-Terry-Luce (BTL) approach and the divide-and-conquer approach. First, the BTL approach was proposed in \cite{bradley1952rank} and is used to rank items based on paired comparisons. Second, in this work, we propose a divide-and-conquer approach, ``CrowDC,'' to reduce the number of tasks required while maintaining accuracy. 


\subsection{BTL Approach}

In the BTL approach, a dataset $D = \{d_1, d_2, \ldots, d_n\}$ containing $n$ items must be ranked, and then the scores for each item must be computed. Specifically, subjects provide paired comparisons $C = \{c_1, c_2, \ldots, c_m\}$, where each comparison $c_i$ is represented as $c_i(id, d_a, d_b, d_c)$. $id$ denotes the subject ID, $d_a$ and $d_b$ represent the two compared items, and $d_c$ indicates the selected item from $d_a$ and $d_b$. Next, the Bradley-Terry (BT) model is employed to calculate the scores $S = \{s_1, s_2, \ldots s_n\}$ from $C$, signifying the scores for each item in $D$. Scores $S$ are then normalized between 0 and 1.

\subsection{Divide-and-Conquer Approach}

Divide-and-conquer is a common algorithmic technique for solving complex problems by dividing them into smaller and simpler subproblems. Thus, in this study, we proposed ``CrowDC,'' utilizing a divide-and-conquer approach to a paired comparison dataset by quantifying the relationship between its comparing results. 


Initially, in the ``Divide'' phase, we divide the dataset $D$ into $g$ groups and obtain $g$ subsets, namely $D.sub_1, \ldots, D.sub_g, D.sub_i = \{sd_{i,j} = d_k | 1 \leq j \leq \frac{n}{g}, d_k \in D\},$ where the number of items in each subset is $\frac{n}{g}$. Second, we collect paired comparisons $C.sub_i$ from subjects for each subset $D.sub_i$. 
Then, we input $C.sub_i$ into the BT model to obtain the individual scores for items in $D.sub_i$, where the scores are specified as within-group scores $S.in_i = \{s.in_j | j \in [1, n]\}$ and $s.in_j$ represents the within-group score for item $d_j$ in the subset.

Next, we assign pivots to each subset to measure the relationship between the estimation scores of each group. We order the items according to scores and choose $p$ items from each subset as pivots. The set of orders is then represented as 

$$ ORD = \{ord_i = min(\lfloor \frac{(\rfrac{n}{g} - 1) \times (i - 1)}{p - 1} \rfloor + 1, \frac{n}{g}) | 1 \leq i \leq p \}. $$

For each subset $D.sub_i$, we first sort $D.sub_i$ based on the scores $S.in_i$ in ascending order, and then we select items based on the previously determined set of orders, $D.piv_i = \{pd_{i,j} = sd_{i,k} | 1 \leq j \leq p, k \in ORD, sd_{i,k} \in D.sub_i\}$, as the pivots. This ensures that the scores in $S.in_i$ of all other items in $D.sub_i$ are between those of the pivots. For example, given a set of 10 items in each subset and a value of 4 for $p$, the set of orders is represented as $ORD = \{1, 4, 7, 10\}$, indicating that the first, fourth, seventh, and tenth items are chosen as pivots from each subset based on their scores. After selecting $D.piv_1, \ldots, D.piv_g$, we combine them as the pivot set $D.piv_{all} = D.piv_1 \cup D.piv_2 \cup \dots \cup D.piv_g$. Likewise, we gather paired comparisons $C.piv$ from subjects, the paired comparison results of every item pair in $D.piv_{all}$. Then, we input $C.piv$ into the BT model to obtain the scores for $D.piv_{all}$, termed out-of-group scores. The value of $S.out = \{s.out_i | i \in [1, n]\}$, where $s.out_i$ denotes the out-of-group score for the item with item number $i$. Algorithm \ref{alg:divide} details the preceding algorithm.

\begin{figure}[h]
  \centering
  \includegraphics[width=.95\linewidth]{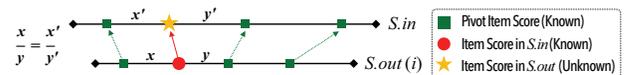}
  \caption{Example for score fitting.}
  \label{fig:ScoreFit}
  \Description{}
\end{figure}

During the ``Conquer'' phase, we compute the final scores $S.fin = \{s.fin_i | 1 \leq i \leq n\}$ for every item, using $S.in_i \ldots S.in_g$ and $S.out$. For pivots, the final scores are their scores in $S.out$. 
For each item $d_j$ in $D.sub_i$, we choose its left closest pivot item $d_l$ and the right closest pivot item $d_r$; then we calculate the final score $s.fin_j$ for $d_j$ utilizing proportionality. Figure \ref{fig:ScoreFit} demonstrates how we fit data from within-group scores to final scores, and Algorithm \ref{alg:conquer} describes the algorithm's procedure. After calculating the quantitative findings for all paired comparing outcomes, we can determine the final scores $S.fin$ for each item in the paired comparison dataset.


\begin{algorithm}
    \caption{Divide Algorithm}\label{alg:divide}
    \begin{algorithmic}[1]
        \Statex \textbf{Input:} dataset, $D$; paired comparisons, $C$; item size, $n$; group count, $g$; pivot count, $p$;
        \Statex \textbf{Output:} within-group score, $S.in$; out-of-group score, $S.out$; sub-dataset, $D.sub$; pivot item, $D.piv$;
    
        \State $D.sub_1 \ldots D.sub_g \gets Divide(D, g)$ \Comment{divide $D$ into $g$ subsets}
        \For{$i \in [1, g]$}
            \State $C.sub_i \gets CollectPC(D.sub_i)$ \Comment{comparisons for i-th subset}
            \State $S.in_i \gets BT(C.sub_i)$
        \EndFor
        \State $D.piv_{all} \gets \{\}$ \Comment{initialize a set for pivots, $D.piv_{all}$}
        \For{$i \in [1, g]$}
            \State $D.piv_i \gets \{\}$
            \State $Sort(D.sub_i, S.in_i)$ \Comment{sort $D.sub_i$ using $S.in_i$}
            \For{$j \in [1, p]$}
                \State $k \gets min(\lfloor \frac{(\rfrac{n}{g} - 1) \times (j - 1)}{p - 1} \rfloor + 1, \frac{n}{g})$
                \State $D.piv_i \gets D.piv_i \cup \{sd_{i,k}\}$ 
            \EndFor
            \State $D.piv_{all} \gets D.piv_{all} \cup D.piv_i$ 
        \EndFor
        \State $C.piv \gets CollectPC(D.piv_{all})$ \Comment{comparisons from pivots}
        \State $S.out \gets BT(C.piv)$
    \end{algorithmic}
\end{algorithm}

\begin{algorithm}
    \begin{algorithmic}[1]

    \caption{Conquer Algorithm}\label{alg:conquer}
        \Statex \textbf{Input:} within-group scores, $S.in$; out-of-group scores, $S.out$; sub-datasets, $D.sub$; pivot items, $D.piv$; group count, $g$; pivot count, $p$;
        \Statex \textbf{Output:} Final scores, $S.fin$;
        \For{$i \in [1, g]$}
            \For{$d_j \in D.sub_i$}
                \For{$k \in [1, p - 1]$}
                    \State $d_l \gets$ $pd_{i,k}$
                    \State $d_r \gets$ $pd_{i,k+1}$
                    \If{$s.in_l \leq s.in_j \leq s.in_r$}
                        \State $s.fin_j \gets s.out_l + \frac{s.in_j - s.in_l}{s.in_r - s.in_l} \times (s.out_r - s.out_l$) 
                    \EndIf
                \EndFor
            \EndFor
        \EndFor 
    \end{algorithmic}
\end{algorithm}

\begin{figure}
     \centering
     \begin{subfigure}[b]{0.235\textwidth}
         \centering
         \includegraphics[width=\textwidth]{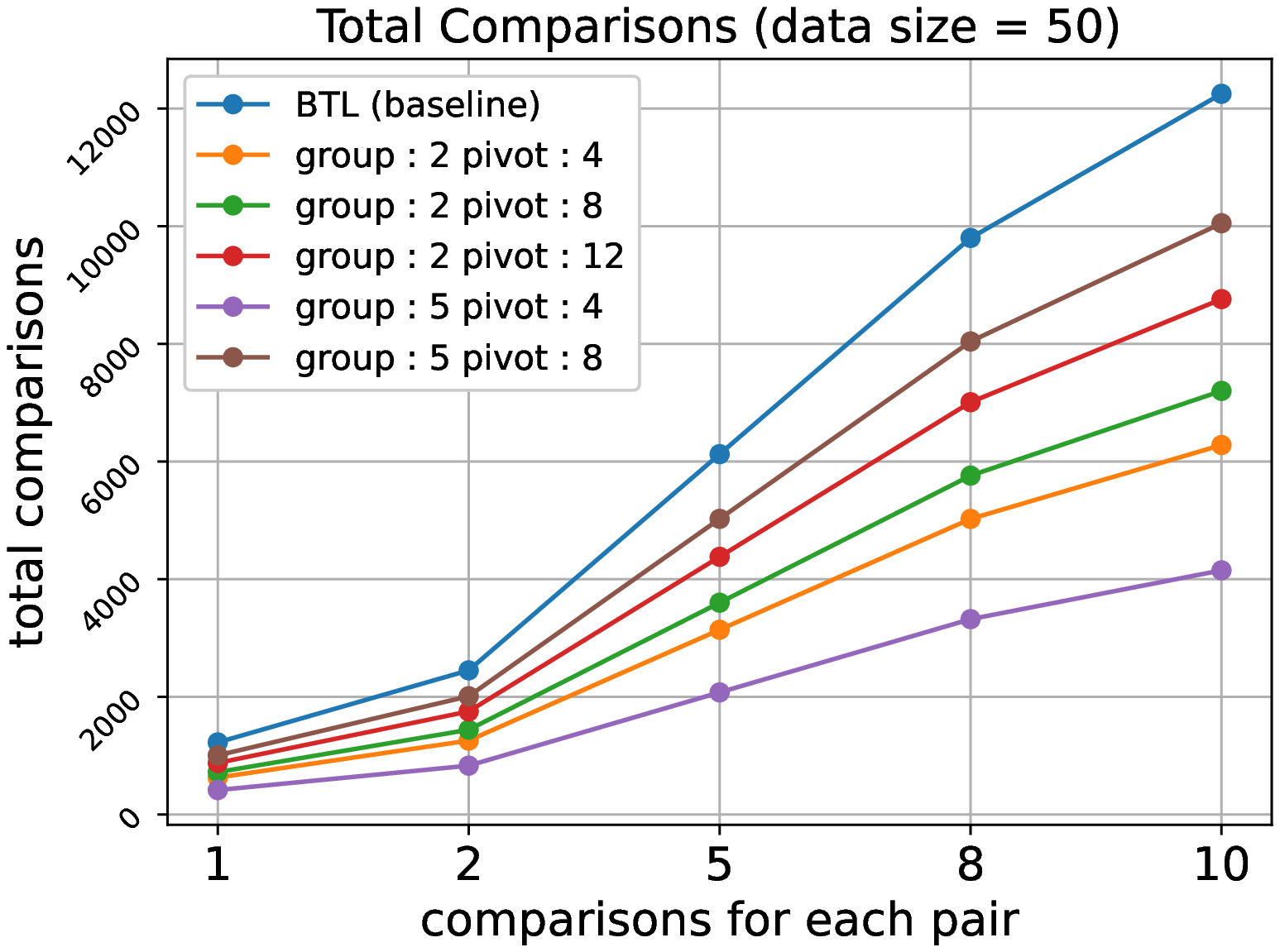}
     \end{subfigure}
     \hfill
     \begin{subfigure}[b]{0.235\textwidth}
         \centering
         \includegraphics[width=\textwidth]{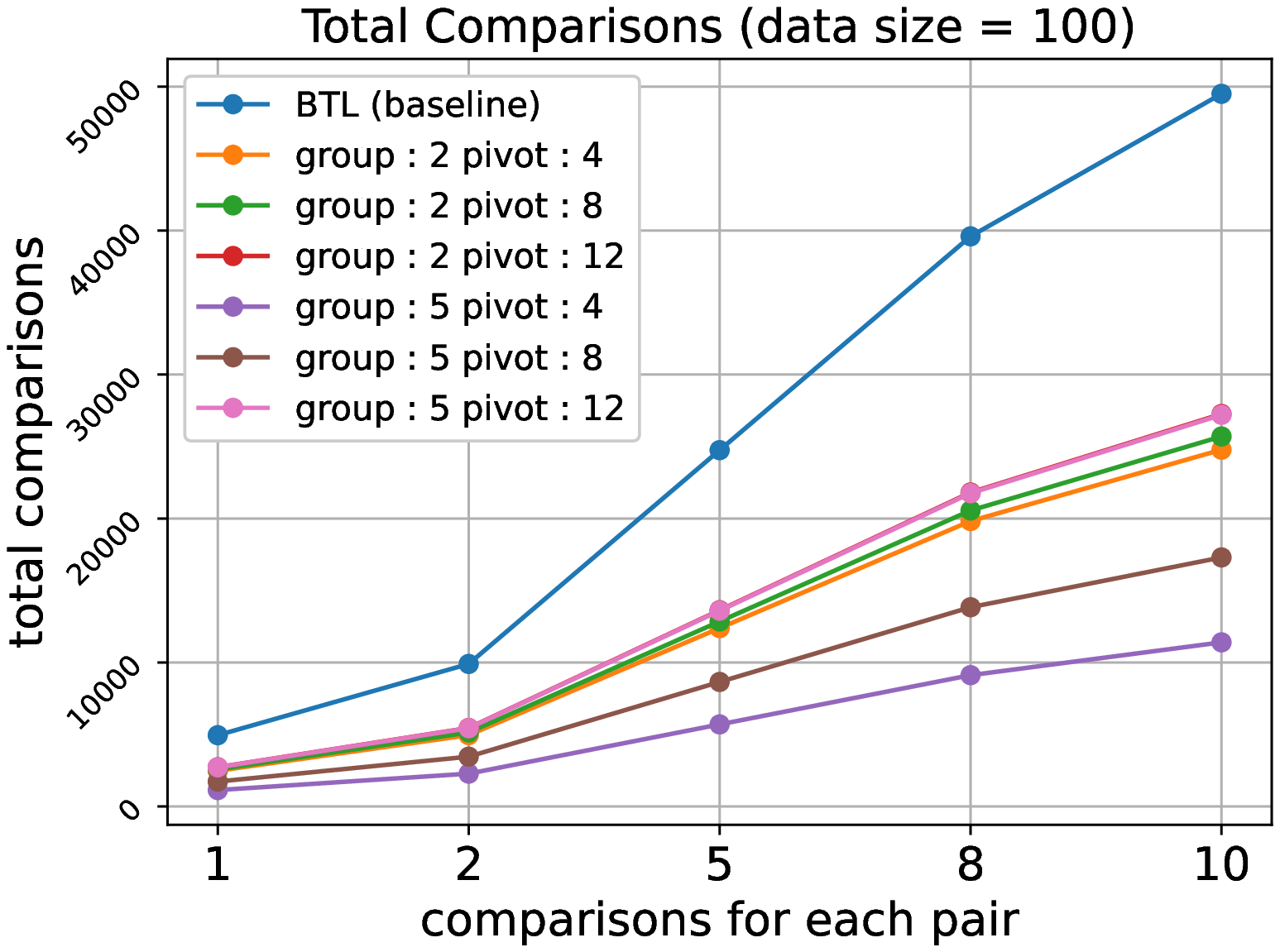}
     \end{subfigure}
     \hfill
     \begin{subfigure}[b]{0.235\textwidth}
         \centering
         \includegraphics[width=\textwidth]{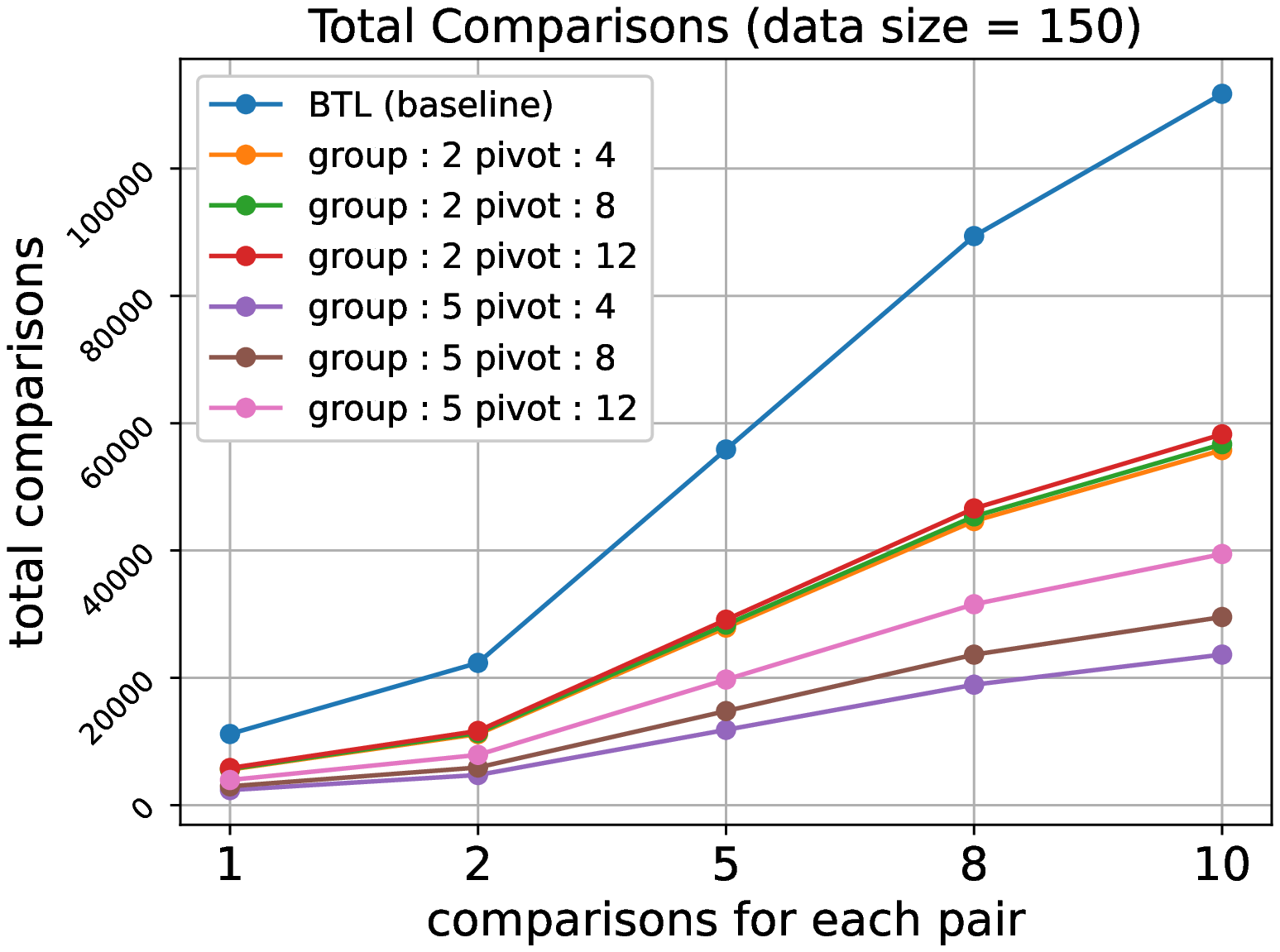}
     \end{subfigure}    
     \hfill
     \begin{subfigure}[b]{0.235\textwidth}
         \centering
         \includegraphics[width=\textwidth]{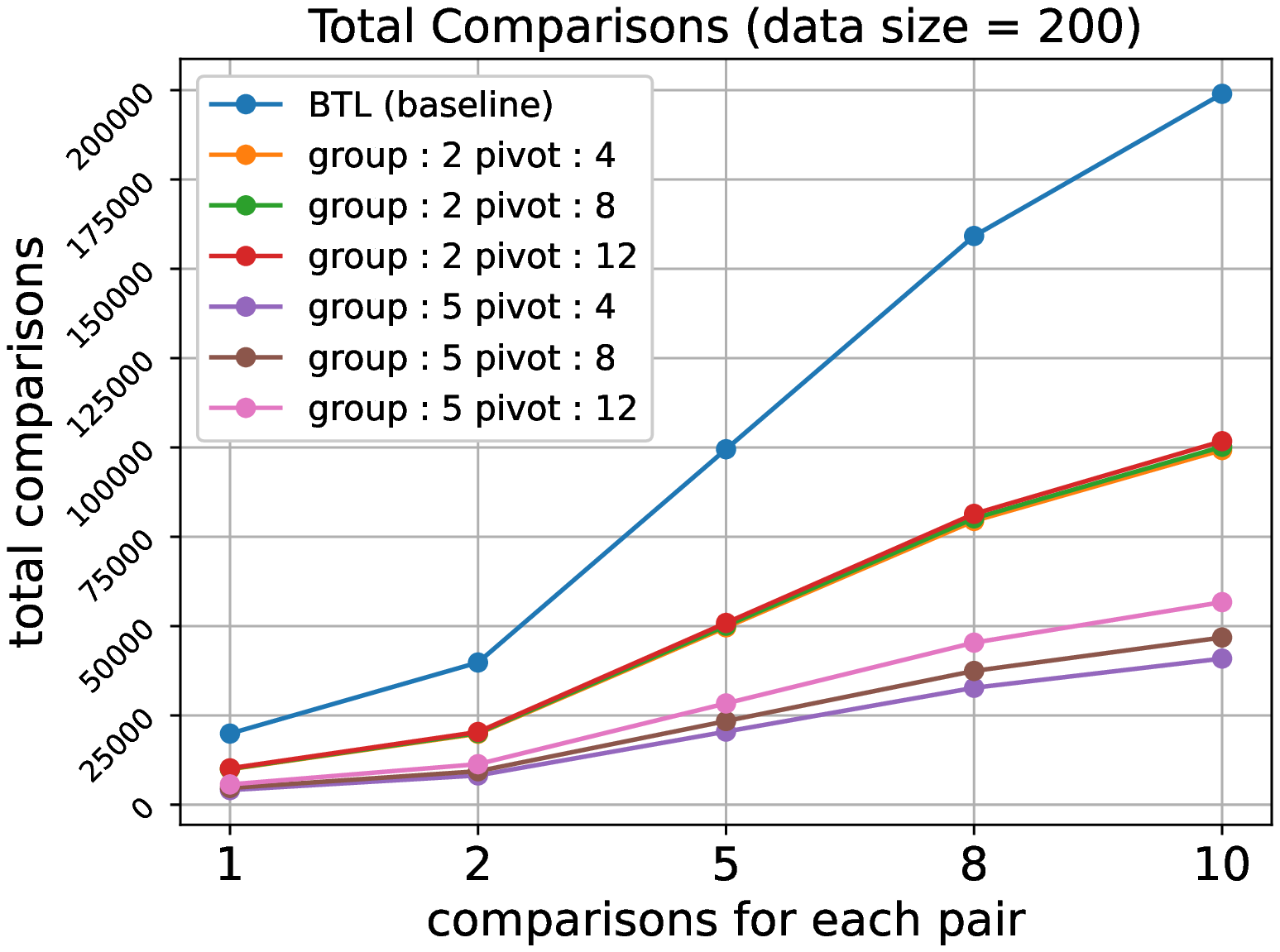}
     \end{subfigure}    
     \caption{Simulation results of ranking items (\# comparisons).}
    \label{fig:Sim_Cmp}
\end{figure}

\begin{figure*}
     \centering
     \begin{subfigure}[b]{0.245\textwidth}
         \centering
         \includegraphics[width=\textwidth]{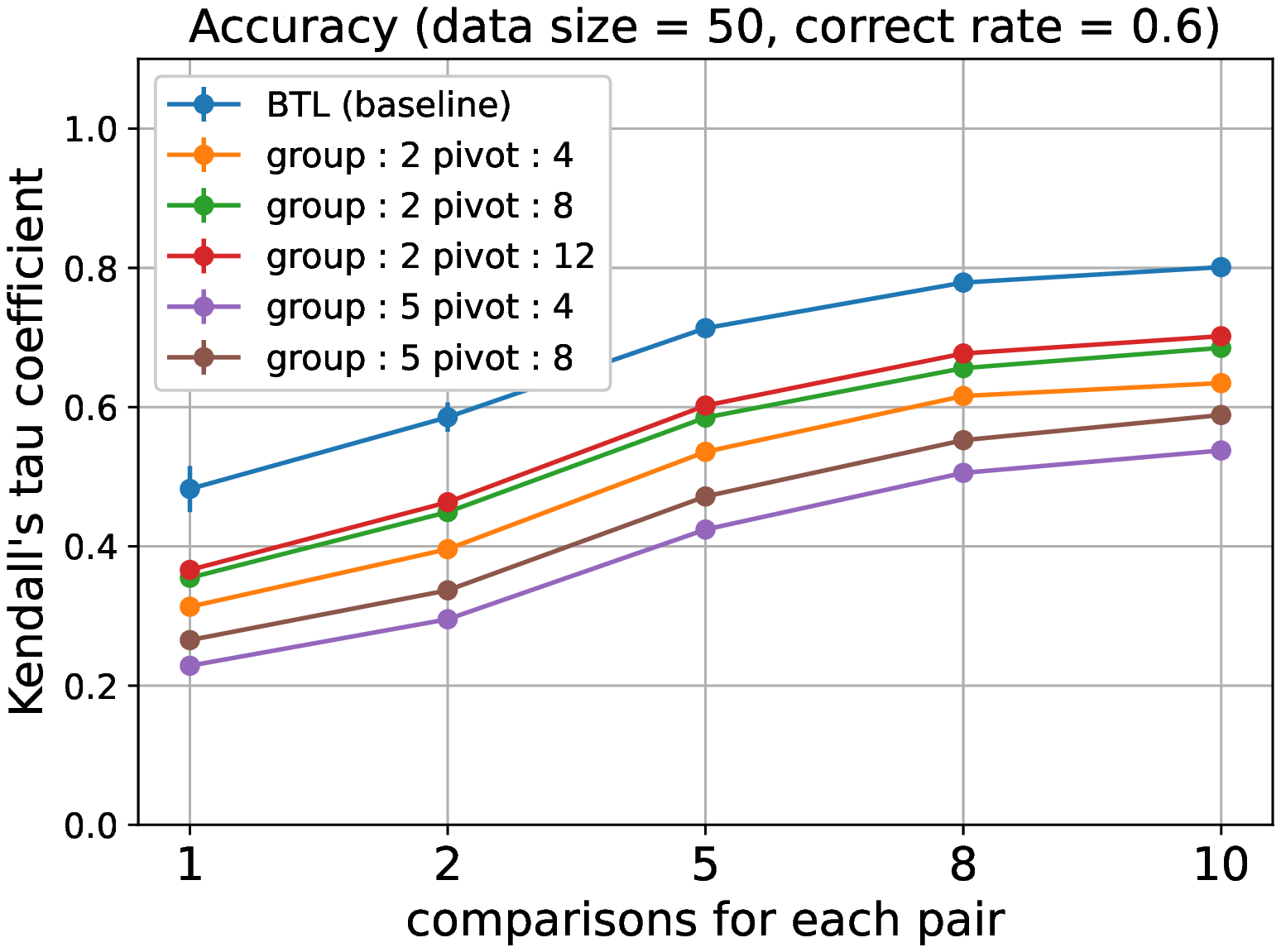}
     \end{subfigure}
     \hfill
     \begin{subfigure}[b]{0.245\textwidth}
         \centering
         \includegraphics[width=\textwidth]{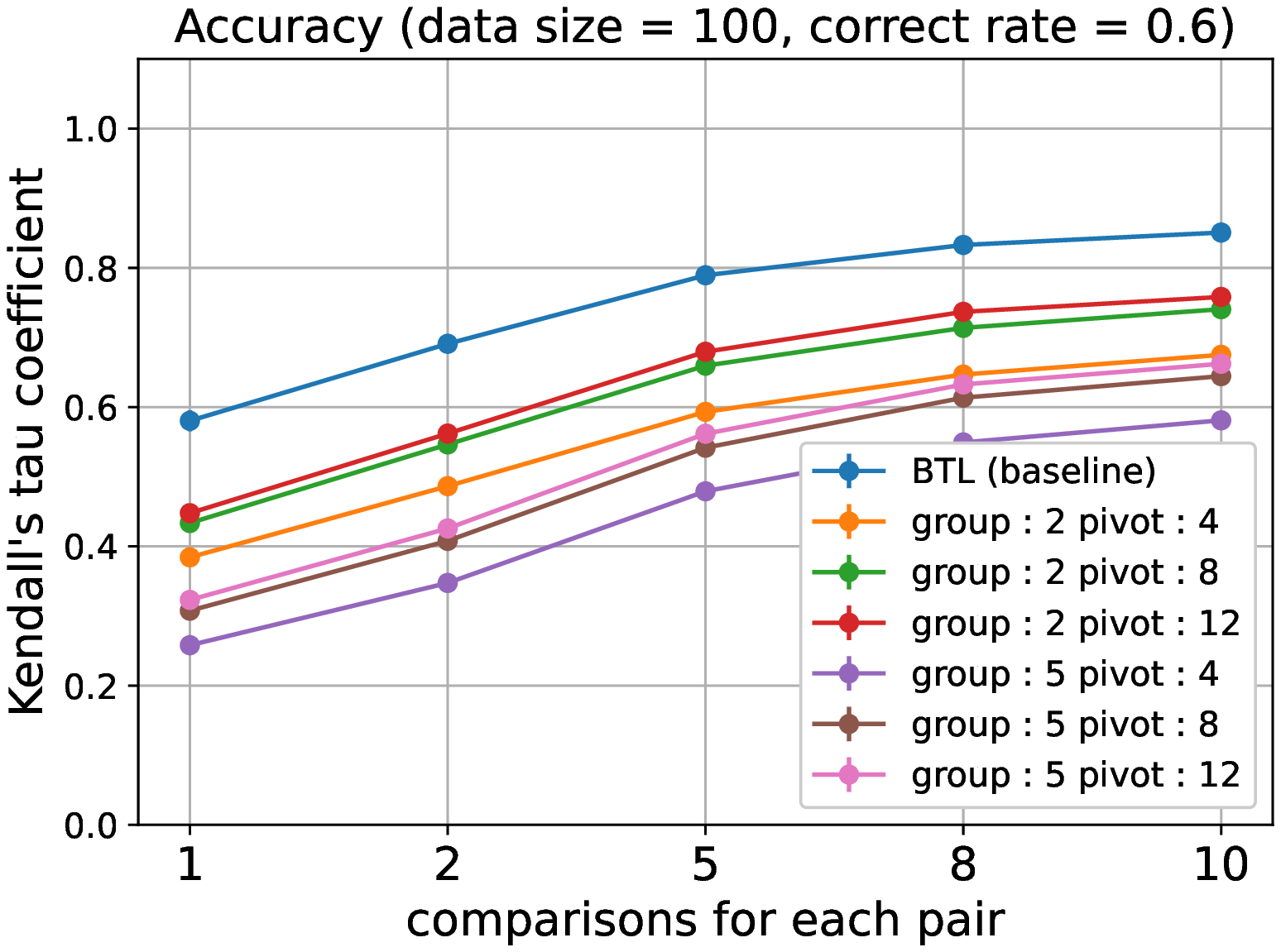}
     \end{subfigure}
     \hfill
     \begin{subfigure}[b]{0.245\textwidth}
         \centering
         \includegraphics[width=\textwidth]{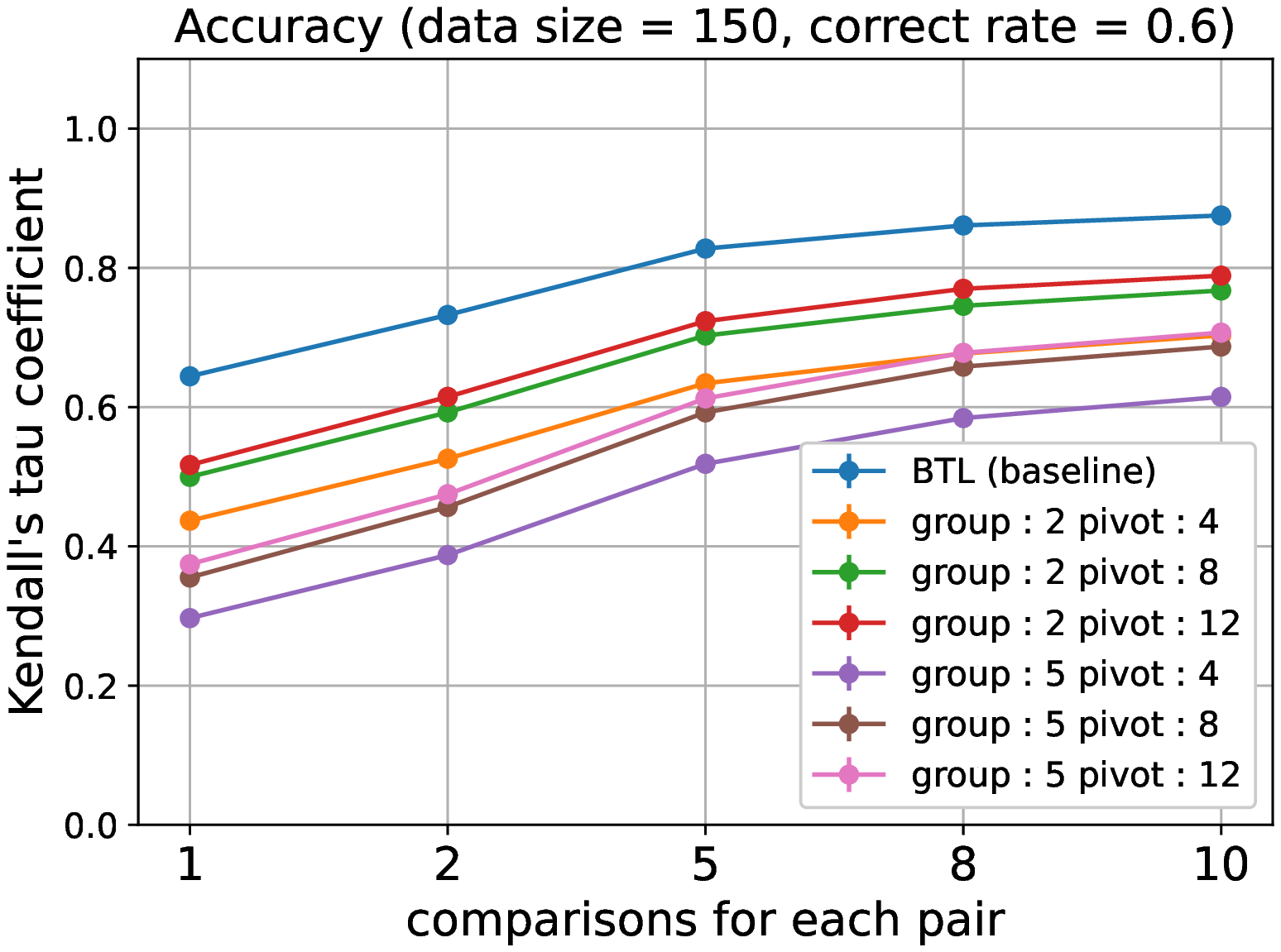}
     \end{subfigure}
     \hfill
     \begin{subfigure}[b]{0.245\textwidth}
         \centering
         \includegraphics[width=\textwidth]{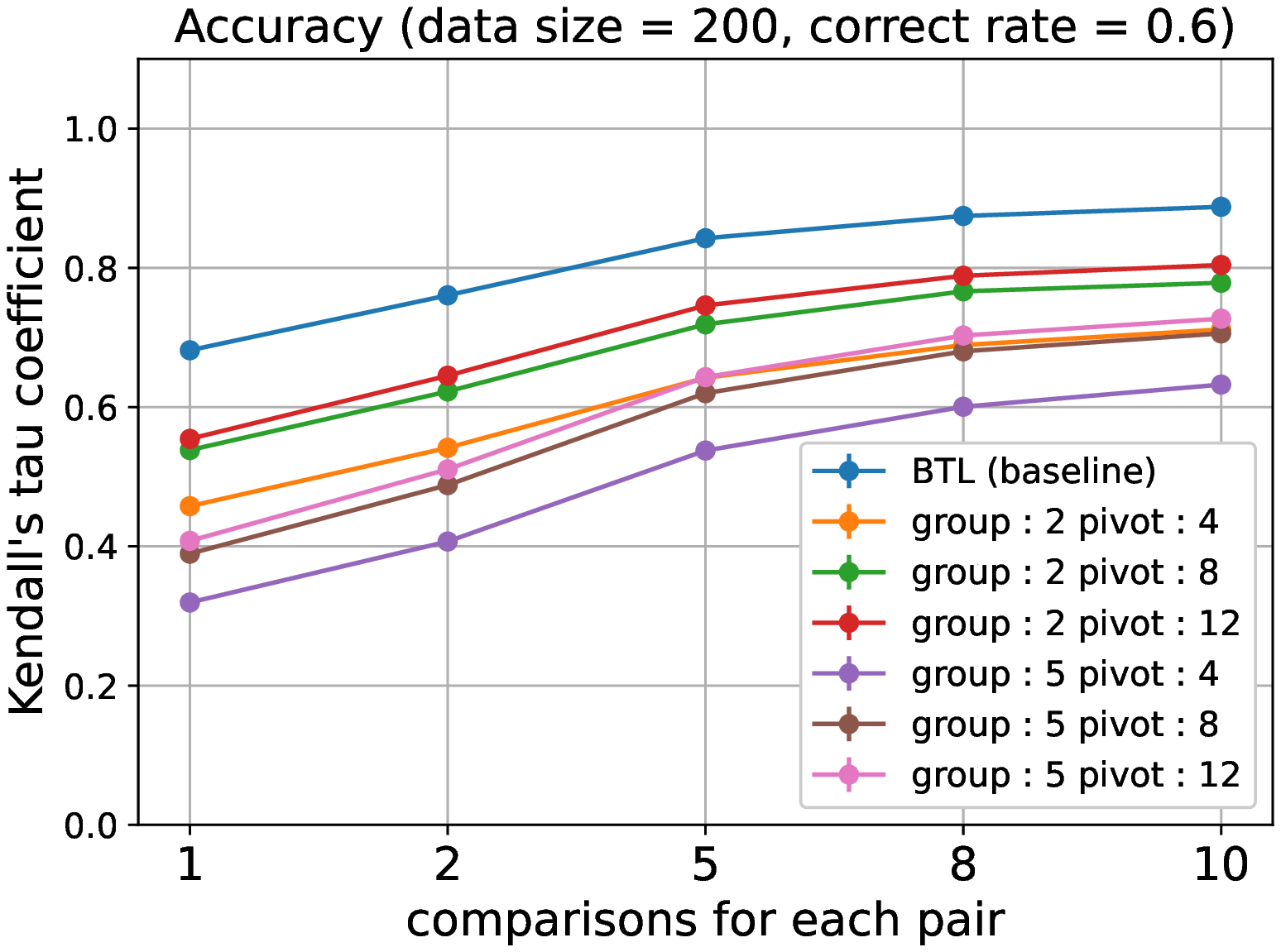}
     \end{subfigure}
     \hfill
     \begin{subfigure}[b]{0.245\textwidth}
         \centering
         \includegraphics[width=\textwidth]{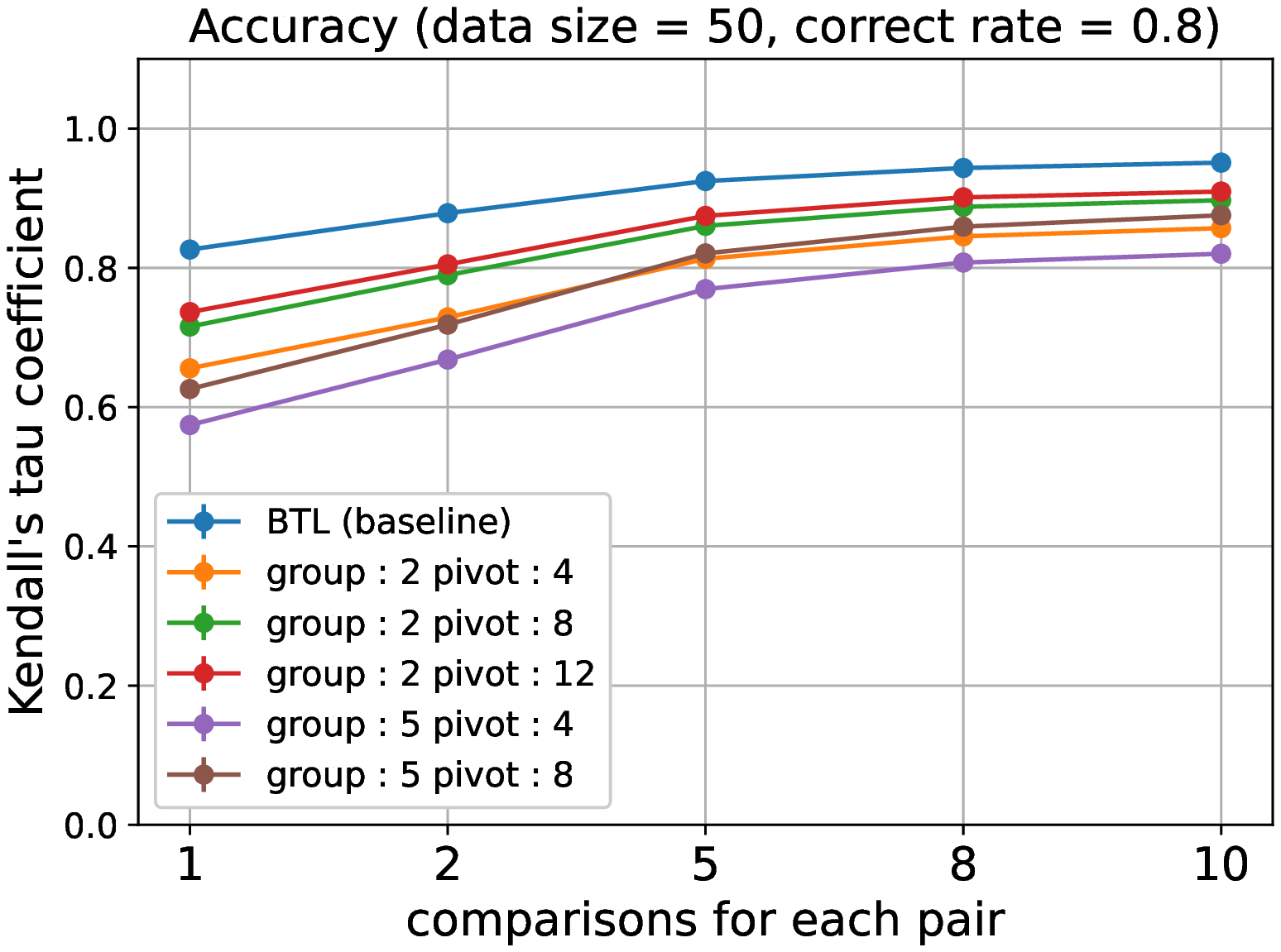}
     \end{subfigure}
     \hfill
     \begin{subfigure}[b]{0.245\textwidth}
         \centering
         \includegraphics[width=\textwidth]{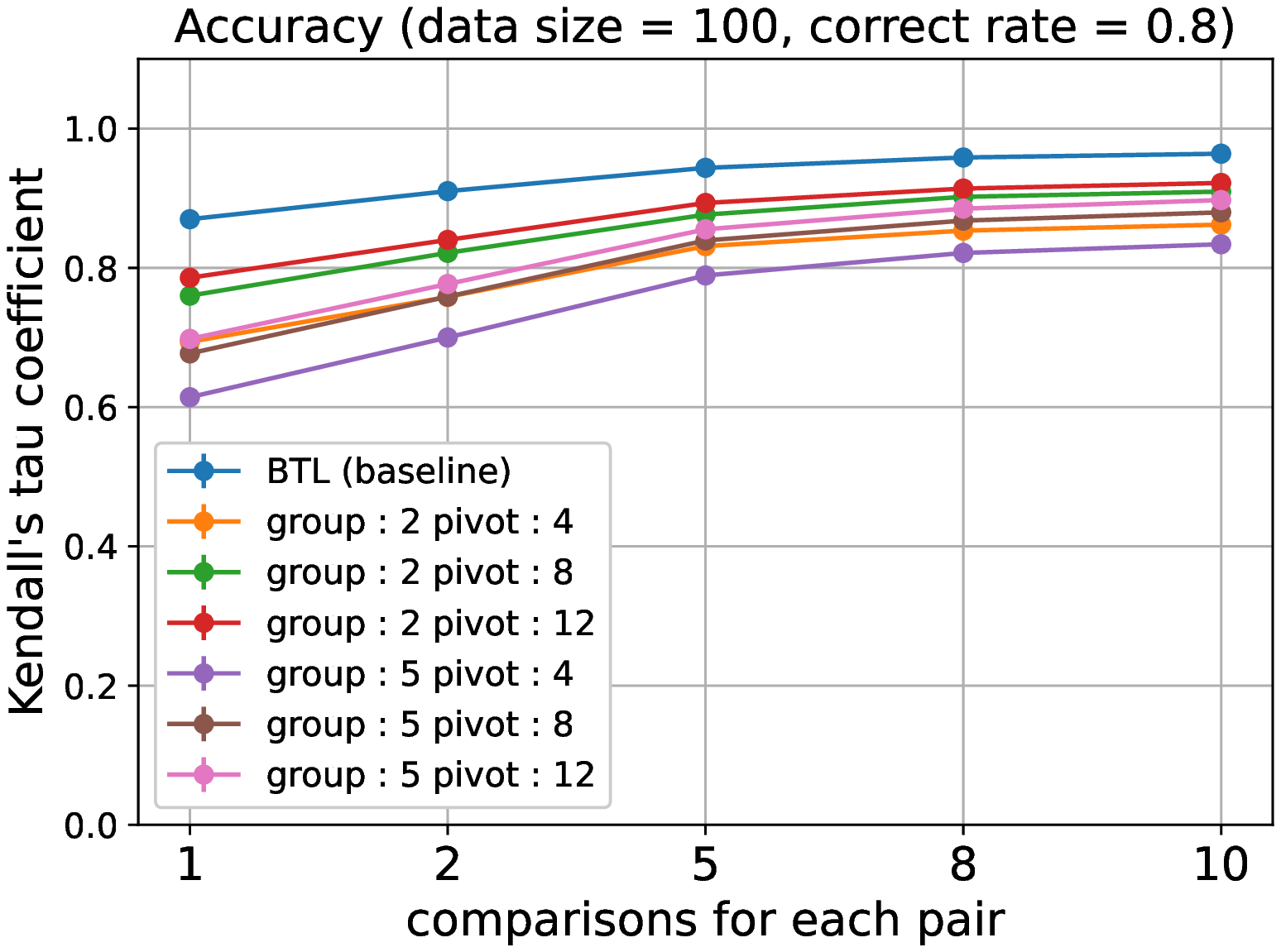}
     \end{subfigure}
     \hfill
     \begin{subfigure}[b]{0.245\textwidth}
         \centering
         \includegraphics[width=\textwidth]{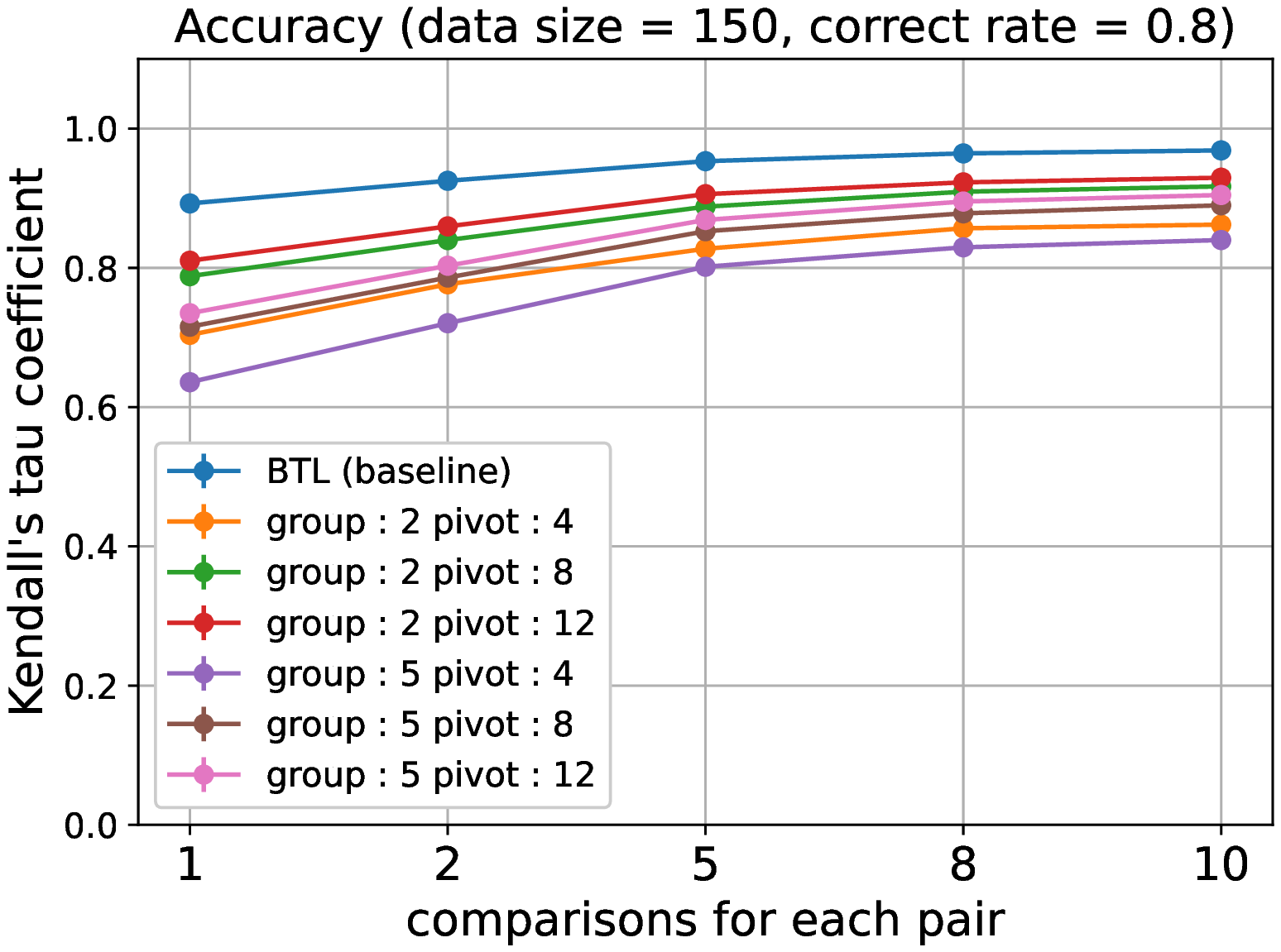}
     \end{subfigure}
     \hfill
     \begin{subfigure}[b]{0.245\textwidth}
         \centering
         \includegraphics[width=\textwidth]{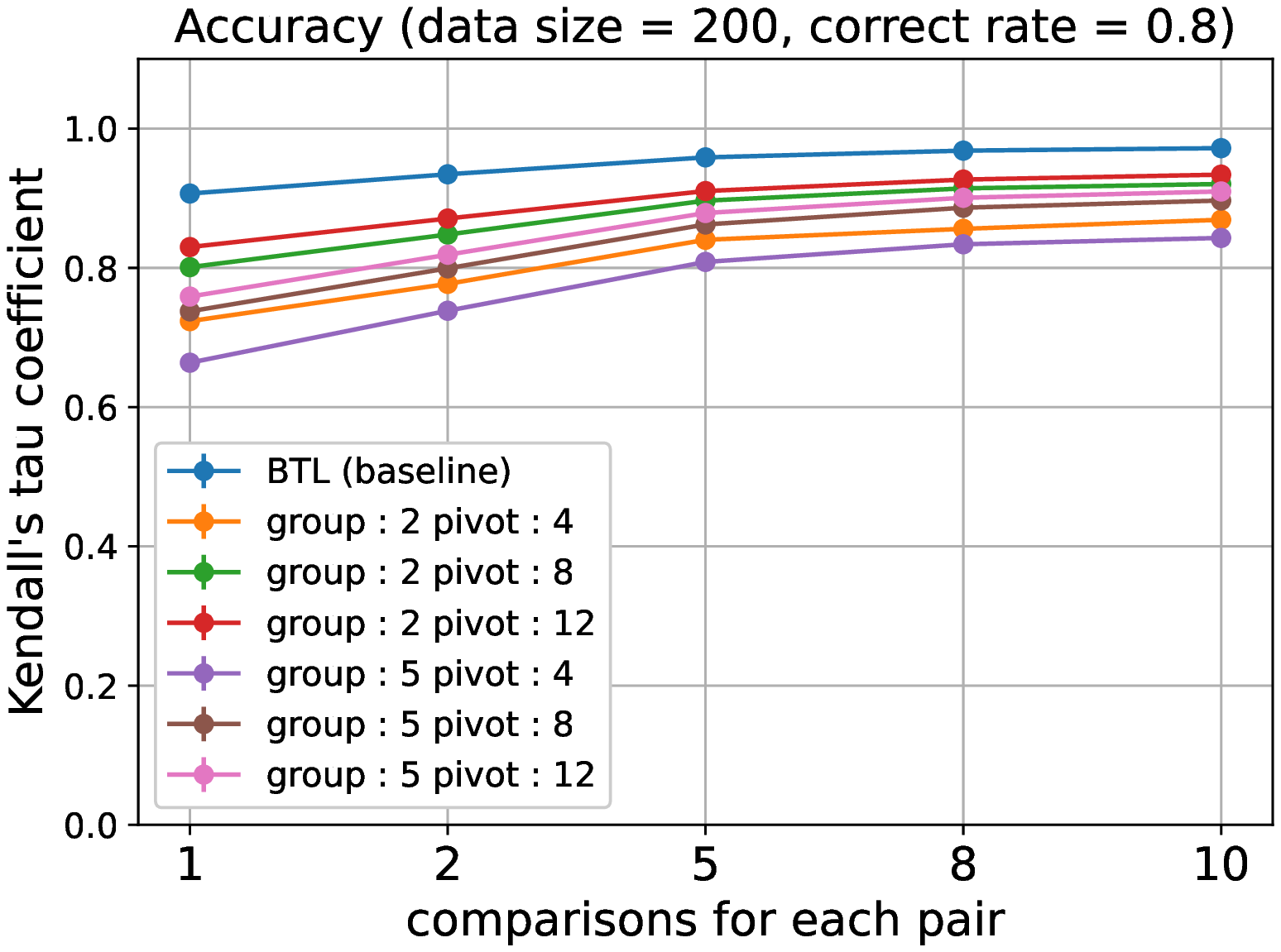}
     \end{subfigure}
    \caption{Simulation results of ranking items (accuracy).}
    \label{fig:Sim_Rank}
\end{figure*}

\section{Simulation and Result}

In this section, we present the simulation results of the BTL approach and CrowDC with various combinations of simulation variables. Then, we apply these two methods to the generated datasets and compare their performance and accuracy.

\begin{table}
  \caption{A summary of simulation variables}
  \label{tab:SimVar}
  \begin{tabular}{ccl}
    \toprule
    Variable description & Variable name & Range \\
    \midrule
    item size                      & $n$ & {[}50, 100, 150, 200{]}        \\
    comparisons for each pair & $t$ & {[}1, 2, 5, 8, 10{]}            \\
    correct rate           & $r$ & {[}0.6, 0.8{]} \\
    group count                    & $g$ & {[}2, 5{]}               \\
    pivot count                    & $p$ & {[}4, 8, 12{]}               \\
    \bottomrule
  \end{tabular}
\end{table}

\subsection{Simulation Setup}


To assess the effectiveness of the BTL approach and CrowDC in ranking situations, we create a dataset with ranking relationships and compare all pairs of items within it.

Table \ref{tab:SimVar} shows the range of our simulation control variables, where $n$ signifies the number of items in the dataset. To initiate our simulation test, let $D = \{d_1, d_2, \ldots, d_n\}$ be the generated dataset including ranking-related items, with $d_b$ preferred over $d_a$ if $b > a$. Next, comparisons for each pair $t$ is the total number of comparison results created by the simulation for each possible pair $(d_a, d_b)$ where $b > a$. In addition, the correct rate $r$ represents the likelihood that the subject selects the correct item. Given a pair $(d_a, d_b)$ where $b > a$, there is a $r$ chance that $d_b$ is the chosen item; otherwise, $d_a$ is selected. We then produce paired comparisons depending on the given item size, comparison times, and correct rate.



We set up different variables for each comparison. When simulating the BTL approach, we only control the first three variables, the item size, comparisons for each pair, and the correct rate. To simulate CrowDC, however, we control the aforementioned three variables in addition to two extra variables, the group count $g$, and the pivot count $p$. Specifically, group count indicates the number of subsets $D.sub_1 \ldots D.sub_g$ that the dataset $D$ is divided into, and pivot count represents the number of item from each subset $D.sub_i$ that are selected for $D.piv_i$ as pivots.


Next, we compare the simulation results of the BTL approach and CrowDC. We represent the total number of comparisons of the BTL approach as $\binom{n}{2} \times c$, and each item will be compared with all other items in the dataset. In contrast, the $\frac{n}{g}$ items inside each subset will be compared and $g \times p$ pivot items. The total comparisons of CrowDC can be denoted as $(\binom{\rfrac{n}{g}}{2} \times g + \binom{g \times p}{2}) \times c$. Additionally, some compared pairs are presented in both the compared pairs for the subsets and the compared pairs for the pivot items, allowing the created comparisons to be shared between the ``Divide'' and ``Pivot'' parts. To measure the accuracy of the estimates produced by CrowDC, we use Kendall's $\tau$ coefficient.
We also compare the results of CrowDC to those of the BTL method (baseline) to assess the level of task reduction and any associated loss in accuracy. 


Furthermore, to make CrowDC calculable, we must consider the scope and limits of certain variables, particularly group count and pivot count, $g$ and $p$. Specifically, $p$ should be greater than $2$ since each subset must contain at least two pivot items (i.e. the item with the highest and lowest within-group scores). For the group count $g$, this variable must be larger than $1$ and less than the size $\frac{n}{3}$, as each subset contains at least three item, including two pivot items and a non-pivot one; otherwise, the ``Conquer'' part of CrowDC will be invalid. In addition, $g$ must be divisible by $d$.




\subsection{Result}

This section provides our simulation results and analyzes the 20 ranking datasets we created for each combination of the simulation variables $(n, t, r, g, p)$. For each dataset, we conduct one simulation for the BTL approach and 20 simulations for CrowDC. The BTL approach and CrowDC differ in the number of simulations because we intend to evaluate the performance of CrowDC under different group divisions. Thus, there are 20 simulation results for each parameter combination for the BTL approach and 400 results for CrowDC. The performance comparisons between CrowDC and the BTL approach are depicted in Figure \ref{fig:Sim_Cmp} (the number of tasks) and Figure \ref{fig:Sim_Rank} (the accuracy measured by Kendall's $\tau$ coefficient). 


We compare the accuracy of the BTL approach and CrowDC based on the simulation results for each variable. In Figure {\ref{fig:Sim_Cmp}}, we observe that CrowDC reduces 40\%-75\% of tasks required for the BTL approach in all scenarios when $n$ is 100 or more; however, when $p$ is close to $\frac{n}{g}$, the number of tasks will be close to the number of tasks for the BTL approach. This suggests that our design would be more beneficial when comparing many items.

From the results shown in Figure {\ref{fig:Sim_Rank}}, the first finding is that CrowDC's accuracy decreases as the number of groups increases. When the group size $g$ is set to 2, the total number of comparisons is reduced by up to 48\% at the cost of 5\% accuracy ratio loss ($n \geq 100, t = 5, r=0.8, p=12$). Though task reduction can be more substantial when the group size increases to 5, the cost of accuracy is also higher, especially when the labels are noisy ($r$ = 0.6). 

The second finding is that increasing the number of pivots $p$ enhances the accuracy ratio. This is because the greater the number of pivots, the more information they may provide for merging items from different groups. When the number of pivots increases from 4 to 8, the accuracy also improves. However, when the number of pivots is increased to 12, the improvement is not as significant. In conclusion, although the accuracy of the BTL approach outperforms that of CrowDC in every scenario, when $n \geq 100, t \geq 5, r=0.8$, and $p=12$, CrowDC saves the cost of the BTL approach by 45-50\% while maintaining 95\% of its accuracy. Though CrowDC doesn't reduce the tasks significantly when $n = 50$, the above parameter settings of $t, r, p$ still yield satisfactory results.

\section{Conclusion and Future Work}

In response to the quadratic growth of comparison tasks, this study demonstrates the possibility of efficiently ranking data collection based on paired comparisons via crowdsourcing at a lower cost, even if the raw data is large enough to burden the aforementioned conventional methods. 
Specifically, we have presented a divide-and-conquer algorithm, ``CrowDC'', to rapidly and inexpensively rank subjective human measurements while ensuring justifiable overall accuracy of estimation results. By splitting the paired comparison dataset into groups 
and merging the quantifying findings from groups, we compute the final score for each item with a significantly reduced workload. In addition, the proposed algorithm outperforms the BTL approach in terms of effectiveness based on simulation results, and the precision doesn't drop much.



Our algorithm addresses ranking tasks using paired comparisons from crowdsourcing efficiently and economically, but it has limits. First, although grouping and merging items may reduce redundant burdens, a large-scale crowdsourced paired comparison experiment should be conducted to verify if our method replicates the simulated findings. Second, the suggested method illuminates dataset processing with ranking-related items, while applications without a ranking relationship may require further tailoring. These issues will be addressed in our future works.




\begin{acks}
This work was supported by the National Science and Technology Council, Taiwan, under the Grant NSTC 111-2222-E-194-003 and MOST 111-2622-E-194-005.
\end{acks}

\bibliographystyle{ACM-Reference-Format}
\bibliography{sample-base}


\begin{thebibliography}{9}


\ifx \showCODEN    \undefined \def \showCODEN     #1{\unskip}     \fi
\ifx \showDOI      \undefined \def \showDOI       #1{#1}\fi
\ifx \showISBNx    \undefined \def \showISBNx     #1{\unskip}     \fi
\ifx \showISBNxiii \undefined \def \showISBNxiii  #1{\unskip}     \fi
\ifx \showISSN     \undefined \def \showISSN      #1{\unskip}     \fi
\ifx \showLCCN     \undefined \def \showLCCN      #1{\unskip}     \fi
\ifx \shownote     \undefined \def \shownote      #1{#1}          \fi
\ifx \showarticletitle \undefined \def \showarticletitle #1{#1}   \fi
\ifx \showURL      \undefined \def \showURL       {\relax}        \fi
\providecommand\bibfield[2]{#2}
\providecommand\bibinfo[2]{#2}
\providecommand\natexlab[1]{#1}
\providecommand\showeprint[2][]{arXiv:#2}

\bibitem[Bradley and Terry(1952)]%
        {bradley1952rank}
\bibfield{author}{\bibinfo{person}{Ralph~Allan Bradley} {and}
  \bibinfo{person}{Milton~E Terry}.} \bibinfo{year}{1952}\natexlab{}.
\newblock \showarticletitle{Rank analysis of incomplete block designs: I. The
  method of paired comparisons}.
\newblock \bibinfo{journal}{\emph{Biometrika}} \bibinfo{volume}{39},
  \bibinfo{number}{3/4} (\bibinfo{year}{1952}), \bibinfo{pages}{324--345}.
\newblock


\bibitem[Draws et~al\mbox{.}(2021)]%
        {draws2021checklist}
\bibfield{author}{\bibinfo{person}{Tim Draws}, \bibinfo{person}{Alisa Rieger},
  \bibinfo{person}{Oana Inel}, \bibinfo{person}{Ujwal Gadiraju}, {and}
  \bibinfo{person}{Nava Tintarev}.} \bibinfo{year}{2021}\natexlab{}.
\newblock \showarticletitle{A checklist to combat cognitive biases in
  crowdsourcing}. In \bibinfo{booktitle}{\emph{Proceedings of the AAAI
  Conference on Human Computation and Crowdsourcing}},
  Vol.~\bibinfo{volume}{9}. \bibinfo{pages}{48--59}.
\newblock


\bibitem[Duan et~al\mbox{.}(2022)]%
        {duan2022influences}
\bibfield{author}{\bibinfo{person}{Xiaoni Duan}, \bibinfo{person}{Chien-Ju Ho},
  {and} \bibinfo{person}{Ming Yin}.} \bibinfo{year}{2022}\natexlab{}.
\newblock \showarticletitle{The influences of task design on crowdsourced
  judgement: A case study of recidivism risk evaluation}. In
  \bibinfo{booktitle}{\emph{Proceedings of the ACM Web Conference 2022}}.
  \bibinfo{pages}{1685--1696}.
\newblock


\bibitem[Gleich and Lim(2011)]%
        {gleich2011rank}
\bibfield{author}{\bibinfo{person}{David~F Gleich} {and}
  \bibinfo{person}{Lek-heng Lim}.} \bibinfo{year}{2011}\natexlab{}.
\newblock \showarticletitle{Rank aggregation via nuclear norm minimization}. In
  \bibinfo{booktitle}{\emph{Proceedings of the 17th ACM SIGKDD international
  conference on Knowledge discovery and data mining}}. \bibinfo{pages}{60--68}.
\newblock


\bibitem[Luce(2012)]%
        {luce2012individual}
\bibfield{author}{\bibinfo{person}{R~Duncan Luce}.}
  \bibinfo{year}{2012}\natexlab{}.
\newblock \bibinfo{booktitle}{\emph{Individual choice behavior: A theoretical
  analysis}}.
\newblock \bibinfo{publisher}{Courier Corporation}.
\newblock


\bibitem[Shah et~al\mbox{.}(2016)]%
        {shah2016stochastically}
\bibfield{author}{\bibinfo{person}{Nihar Shah}, \bibinfo{person}{Sivaraman
  Balakrishnan}, \bibinfo{person}{Aditya Guntuboyina}, {and}
  \bibinfo{person}{Martin Wainwright}.} \bibinfo{year}{2016}\natexlab{}.
\newblock \showarticletitle{Stochastically transitive models for pairwise
  comparisons: Statistical and computational issues}. In
  \bibinfo{booktitle}{\emph{International Conference on Machine Learning}}.
  PMLR, \bibinfo{pages}{11--20}.
\newblock


\bibitem[Wu et~al\mbox{.}(2021)]%
        {wu2021learning}
\bibfield{author}{\bibinfo{person}{Aoyu Wu}, \bibinfo{person}{Liwenhan Xie},
  \bibinfo{person}{Bongshin Lee}, \bibinfo{person}{Yun Wang},
  \bibinfo{person}{Weiwei Cui}, {and} \bibinfo{person}{Huamin Qu}.}
  \bibinfo{year}{2021}\natexlab{}.
\newblock \showarticletitle{Learning to automate chart layout configurations
  using crowdsourced paired comparison}. In
  \bibinfo{booktitle}{\emph{Proceedings of the 2021 CHI Conference on Human
  Factors in Computing Systems}}. \bibinfo{pages}{1--13}.
\newblock


\bibitem[Wu et~al\mbox{.}(2013)]%
        {wu2013crowdsourcing}
\bibfield{author}{\bibinfo{person}{Chen-Chi Wu}, \bibinfo{person}{Kuan-Ta
  Chen}, \bibinfo{person}{Yu-Chun Chang}, {and} \bibinfo{person}{Chin-Laung
  Lei}.} \bibinfo{year}{2013}\natexlab{}.
\newblock \showarticletitle{Crowdsourcing multimedia QoE evaluation: A trusted
  framework}.
\newblock \bibinfo{journal}{\emph{IEEE transactions on multimedia}}
  \bibinfo{volume}{15}, \bibinfo{number}{5} (\bibinfo{year}{2013}),
  \bibinfo{pages}{1121--1137}.
\newblock


\bibitem[Xu et~al\mbox{.}(2012)]%
        {xu2012online}
\bibfield{author}{\bibinfo{person}{Qianqian Xu}, \bibinfo{person}{Qingming
  Huang}, {and} \bibinfo{person}{Yuan Yao}.} \bibinfo{year}{2012}\natexlab{}.
\newblock \showarticletitle{Online crowdsourcing subjective image quality
  assessment}. In \bibinfo{booktitle}{\emph{Proceedings of the 20th ACM
  international conference on Multimedia}}. \bibinfo{pages}{359--368}.
\newblock


\end{thebibliography}










\end{document}